\documentclass[twocolumn]{svjour3}         
\smartqed  
\usepackage{graphicx}
\usepackage{amssymb,amsmath,float}
\begin{document}

\title{The dune size distribution and scaling relations of barchan dune fields
}


\author{Orencio Dur\'an 	\and
		Veit Schw\"ammle	\and
		Pedro G. Lind	\and
		Hans J. Herrmann
}


\institute{O. Dur\'an \at
              Institute for Computational Physics, Stuttgart University,
              70569 Stuttgart, Germany \\
              \email{o.duran@utwente.nl}           \\
              \emph{Present address:} of O. Dur\'an  \at 
              Multi Scale Mechanics (MSM), Twente University, 7500 AE Enschede,
              The Netherlands \\
              Tel.: +31 53 489 2694 \\
           \and
           V. Schw\"ammle \at
              Centro Brasileiro de Pesquisas F\'isicas,
             22290-180 Urca - Rio de Janeiro - RJ, Brazil \\
           \and
           P.~G.~Lind \at 
              Centro de Física Teórica e Computacional, Universidade de Lisboa
              Avenida Professor Gama Pinto 2, 1649-003 Lisboa, Portugal \\
           \and
           H.~J.~Herrmann \at 
              Computational Physics, IfB, HIF E12, ETH H\"onggerberg,
             CH-8093 Z\"urich, Switzerland \\
              Departamento de F\'{\i}sica, Universidade Federal do Cear\'a,
             60451-970 Fortaleza, Cear\'a, Brazil
}

\date{Received: date / Accepted: date}

\maketitle

\begin{abstract}
Barchan dunes emerge as a collective phenomena involving the generation of thousands of them in  so called barchan dune fields. By measuring the size and position of dunes in Moroccan barchan dune fields, we find that these dunes tend to distribute uniformly in space and follow an unique size distribution function. We introduce an analytical mean-field approach to show that this empirical size distribution emerges from the interplay of dune
collisions and sand flux balance, the two simplest mechanisms for size selection. The analytical model also predicts a scaling relation between the fundamental macroscopic properties characterizing a dune field, namely the inter-dune spacing and the first and second moments of the dune size distribution.
\end{abstract}

\section{Introduction}
\label{intro}

\begin{figure*}[htb]
\includegraphics[width=0.82\textwidth]{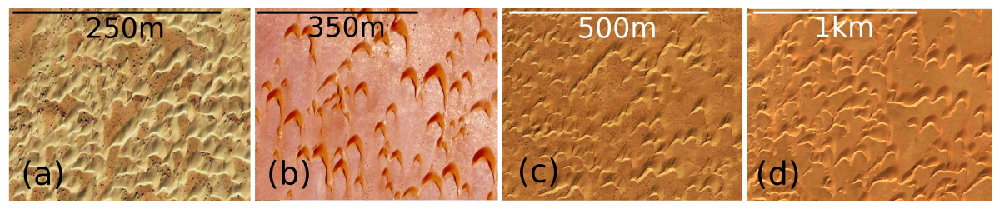}
\includegraphics[width=0.175\textwidth]{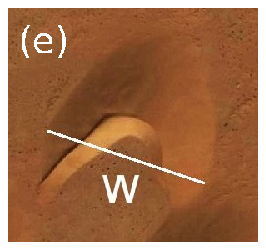}
\includegraphics[width=1.0\textwidth]{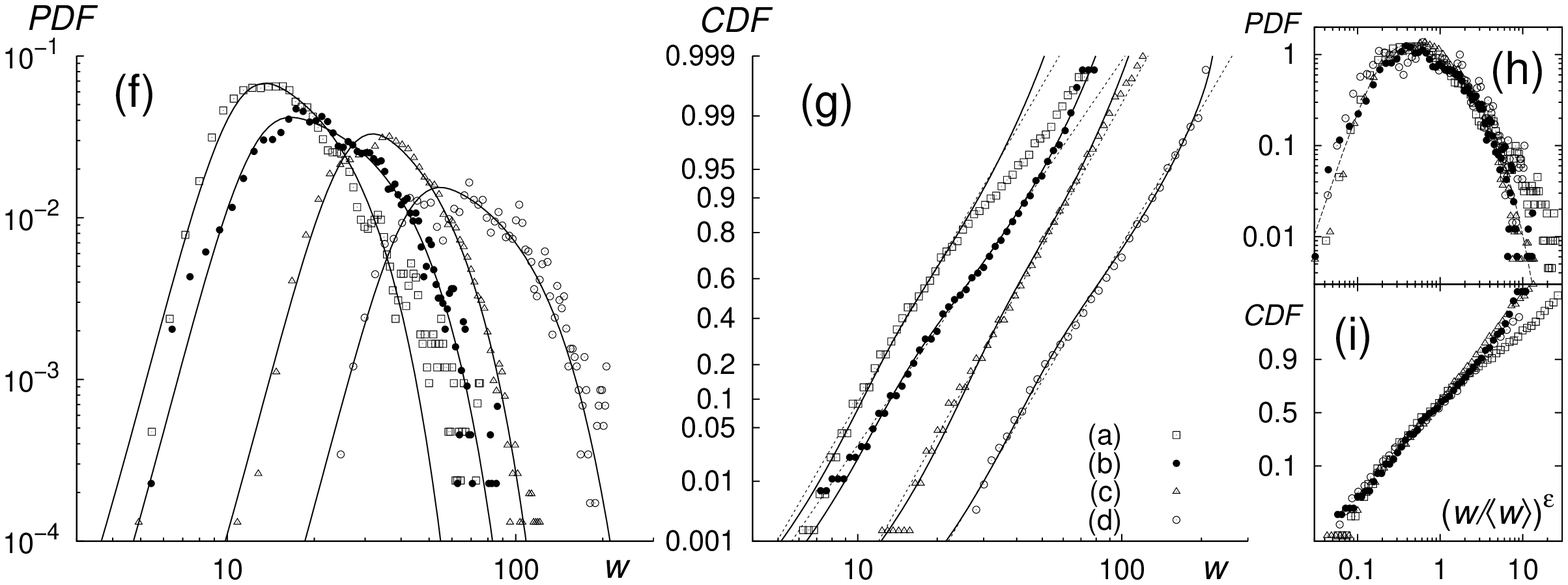}
\caption{\protect (Color online)
{\bf (a-d)} Details of barchan dune fields in Morocco, Western Sahara.
The number of measured dunes is 1295 ({\bf a}), 1113 ({\bf b}), 1947
({\bf c}) and 1630 ({\bf d}), covering areas of $\sim 3, 7, 12$ and $60$
km$^2$ and with average dune sizes of 17m, 27m, 42m and 86m  respectively.
The size of a barchan dune is characterized by its width $w$ {\bf (e)}.
In all pictures, the North points up. Images provided by GoogleEarth.
{\bf (f)} Probability density function (PDF) of the dune size for the
measured Moroccan dune fields (symbols) and the best fit using the
analytical solution (solid lines) given by Eq.~(\ref{boltzmann}).
{\bf (g)} Cumulative distribution function (CDF) with the analytical
solution (solid lines) and the log-normal straight-line for reference
(dashed lines). The relative broadness $S/\langle w\rangle$ is given by
the inverse of the slope of the CDF.
After rescaling the dune sizes as $(w/\langle
w\rangle)^\epsilon$ with $\epsilon = 2.9\sqrt[3]{\langle
w\rangle/\langle L\rangle}$, all PDFs and CDFs in {\bf (h)} and
{\bf (i)} respectively collapse, uncovering a scale invariance
between the size distributions of different barchan dune fields.
In {\bf (h)} the analytical distribution (see text)
is also shown (dashed line), as an eye-guide.
}
\label{fig1}
\end{figure*}

A first glance over an extensive desert area shows not only that dunes are  ubiquitous and present well-selected shapes, but also that they typically  emerge in groups with a very well defined characteristic dune size and inter-dune spacing, forming fields of up to several thousands of dunes  (see Figs.~\ref{fig1}a-d). These observations naturally rise questions concerning the way dunes distribute throughout the deserts. What are the mechanisms of the size selection process behind such uniformity?  Do the size distributions of such dune fields follow a simple unique function  or do they depend on the local conditions?

Sand dunes have been intensively studied in the last years.
It is now well-understood what are the fundamental laws underlying the emergence of one single barchan dune and what mechanisms maintain its shape  while moving~\cite{kroy02,andreotti02}. For instance, barchans occur in areas with unidirectional wind and low sand availability. The influence  of the geographical constrains and the external physical conditions~\cite{andreotti02,hersen04}, of the dune-dune interactions~\cite{schwaemmle03,duran05,hersen05} and even of the emergence of vegetation covers~\cite{duran06} in the dynamics and morphology of single dunes were quite well-established with the help of dune models~\cite{kroy02,veit3D,duran06}. There are also a few studies of entire dune fields~\cite{LSHK02,hersen04b,ewing06,elbelrhiti06}, but a simple theoretical understanding of the size selection process within dune fields has still not been achieved.

In this work, we present a first answer to this problem.
First, we show that, while a single dune is suitably characterized by
its width $w$~\cite{andreotti02,elbelrhiti06}, an entire dune field contains
dunes with different sizes following a unique distribution
(Fig.~\ref{fig1}f-g).
Consequently, the corresponding average $\langle w \rangle$ and standard
deviation $S=\sqrt{\langle w^2\rangle-\langle w\rangle^2}$ are suitable
properties to characterize the field.
Additionally to these two quantities, we show that the inter-dune spacing $L$
is also a property with
characteristic
values $\langle L\rangle$ and
therefore also able to characterize the field. Second, using numerical
simulations, we show that collisions between dunes play a crucial role in
the selection of a characteristic dune size.
Finally, from a mean-field approach that couples the effect of dune collision
with that of sand flux balance, we derive the size distribution function
(shown in Fig.~\ref{fig1}f-g) and a scaling relation between the three
properties of the field,  $\langle L \rangle$, $\langle w \rangle$ and $S$.

\section{Measurements of the dune size and inter-dune spacing distribution}
\label{sec:1}

We start by measuring the width (Fig.\ref{fig1}e) of more than $5000$
dunes composing four dune fields located in the Western Sahara
(Figs.~\ref{fig1}a-d).
For all fields, the dune width distribution exhibits a unique
function, apart small deviations at the extremes ($w\lesssim 10$m
and $w\gtrsim 0.5$km\cite{elbelrhiti06}), as shown
in Fig.\ref{fig1}f-g. This function will be derived later.
The relative broadness $S/\langle w\rangle$ scales with the relative inter-dune spacing as
$\sqrt[3]{\langle L\rangle/\langle w\rangle}$, as shown in
Figs.~\ref{fig1}h-i. This scaling law relates
the spatial distribution of dunes and their size distribution and
will also be deduced later as a result of the size selection model we
propose.
Therefore, the mechanisms leading to such distributions should not depend
on the absolute size and inter-dune spacing of the dunes involved.
Instead, they should depend on the {\it relative} dune size and spacing.
In other words, they should be scale invariant.

\begin{figure*}[htb]
\begin{center}
\includegraphics[width=0.65\textwidth]{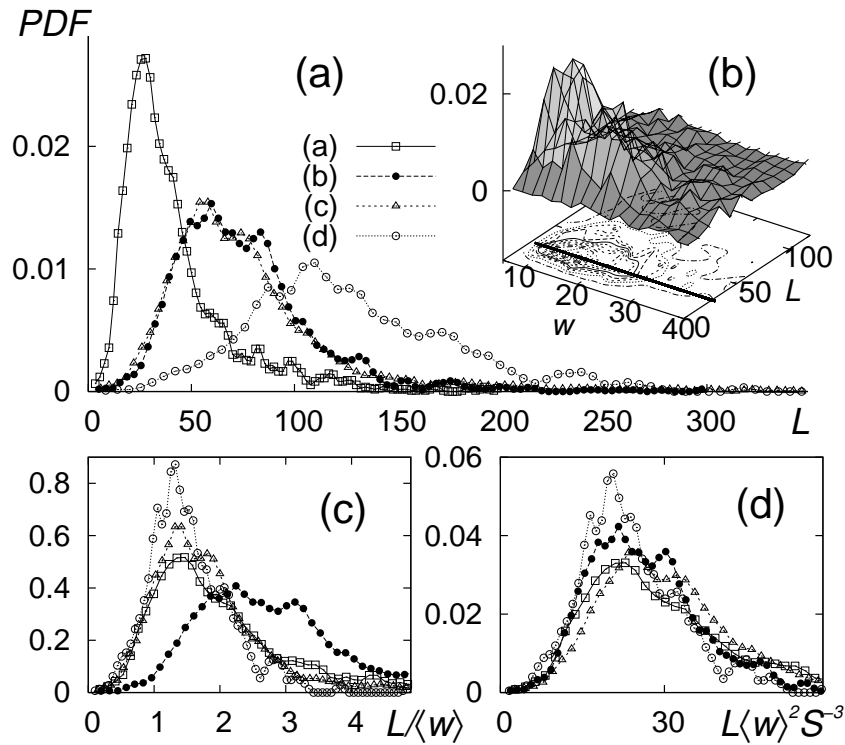}
\end{center}
\caption{\protect {\bf (a)} The PDF of the inter-dune spacing
clearly shows a characteristic value $\langle L \rangle$ for each dune
field depicted in Fig.\ref{fig1}(a-d). The inter-dune spacing $L$
around a given dune is defined as the square root of the empty area of
a polygon formed by the centers of the nearest dunes, one in each
quadrant of the Cartesian frame reference centered at the dune. {\bf
(b)} PDF of $L$ as a function of the dune width $w$ for the first
dune field {\bf a}. From the contour plot, one defines the
characteristic inter-dune spacing $\langle L \rangle$ (solid line),
taken as the average over the highest frequency region, which is
independent of $w$ (see text). {\bf (c)} After rescaling $L$ by the
mean dune size $\langle w\rangle$, not all PDFs peak in the
same relative inter-dune spacing $\langle L \rangle/\langle
w\rangle$. {\bf (d)} However, the curves collapse after rescaling $L$
by the expression $S^3/\langle w\rangle^2$, where $S$
represents the standard deviation of the size distribution (see text).}
\label{fig2}
\end{figure*}

In barchan dune fields, the sand flux balance on single dunes leads
to an instability in the dune size~\cite{hersen05}. Dunes nucleate as
a  consequence of the sand accumulation along the field and no
characteristic  size emerges \cite{hersen05,hersen04b}. However, the
peaked size distributions in Fig.~\ref{fig1}f are found in dune fields
where sand flux balance is not the only process mediating the size of
dunes. Since barchan dunes move over the field with velocities that
strongly depend on the size ($v \propto 1/w$) \cite{kroy02}, collisions are
ubiquitous in such fields, turning out to be another relevant process
for dune size alterations~\cite{hersen05,elbelrhiti06}.
Therefore, the dune size distribution should be determined by the competition between the balance of sand flux on a single dune and the collisions between
neighboring dunes~\cite{duran05,hersen05,hersen04b}.
Recently a third size selection mechanism was discovered,
that involves the calving of large dunes due to wind fluctuations
\cite{elbelrhiti06,elbelrhiti08}. This is a complex scale dependent process,
relevant for fields with large dunes and should not be responsible for
the distributions here addressed. Therefore, we will not consider it.

Two important aspects must now be addressed to proceed further. First,
we notice that in the absence of motion, i.e.~in static fields where
no collisions can occur, one dune grows only if its neighboring dunes
shrink~\cite{kocurek92}, due to sand flux balance and mass
conservation. Consequently, if collisions
do not occur
the inter-dune spacing $L$ between neighboring dunes would scale with the
dune size. Our empirical data however, shows a rather different
behavior.
For each dune field in Figs.~\ref{fig1}a-d, the inter-dune
spacing $L$ between each dune and its neighbors as a function of the
width $w$, distributes parallel to the $w$-axis (Fig.~\ref{fig2}b).
Thus, contrary to the situation without collisions,
here the inter-dune spacing can be taken as its characteristic
value, say $\langle L\rangle$, as shown in Fig.~\ref{fig2}a and
\ref{fig2}b.
This empirical result is a clear sign of a richer
internal dynamics in dune fields where collisions play an important
role. Indeed, due to collisions, small dunes are continuously emerging
from larger ones \cite{elbelrhiti06,elbelrhiti08}, destroying any simple
correlation between dune size and inter-dune spacing, and
therefore leading to the observed spatial uniformity.

The second aspect is that
$\langle L\rangle$ does
not follow a simple scaling with the average dune size $\langle
w\rangle$ of the field (Fig.~\ref{fig2}c) but a more complex one (as
$S^3/\langle w\rangle^2$, Fig.~\ref{fig2}d) which also involves the
standard deviation $S$ of the size distribution. As will be shown later, this
scaling naturally emerges from the coupling of binary dune collisions
and sand flux balance in the size selection process.

\section{Binary collision dynamics}
\label{sec:2}

\begin{figure}[htb]
\begin{center}
\includegraphics[width=0.4\textwidth]{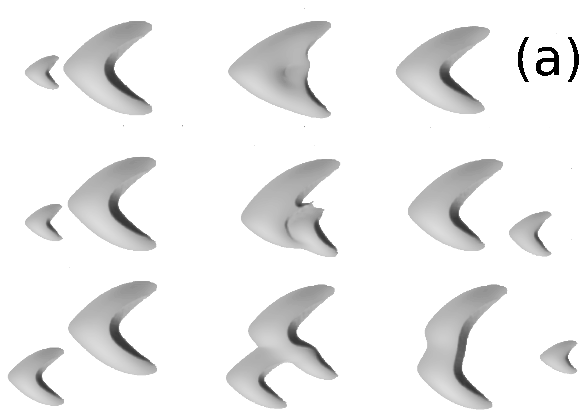}

\includegraphics[width=0.4\textwidth]{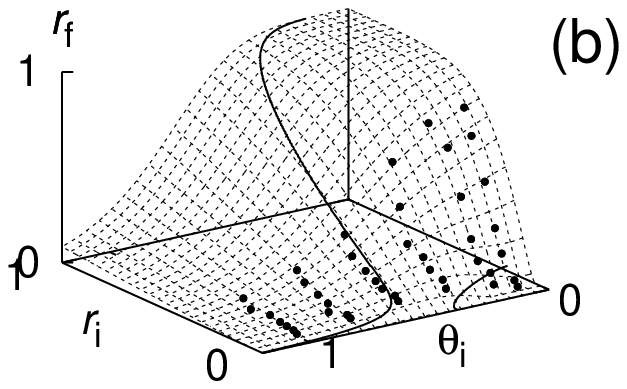}
\end{center}
\caption{\protect
{\bf (a)} Different outcomes of simulated binary collisions, from coalescence (when both dunes merge) on top, to a situation where the volume of the smaller dune increases (decreases) after the collision (middle and bottom sections, respectively). {\bf (b)} Collision rule for binary collisions that conserve the number of dunes. Dots represent numerical simulations and dashed lines the corresponding surface fit $r_f(r_i,\theta_i)$. The curve $r_f(r_i, \theta_i)=r_i$ (solid line) separates two regimes: one with $r_f > r_i$, where collisions redistribute sand and another with $r_f < r_i$ due to accumulation of sand.
}
\label{fig3}
\end{figure}

With these two empirical findings we proceed by studying
the effect of collisions alone in the size
distribution of a dune field. Using an established dune
model~\cite{kroy02,veit3D,duran05}, we simulate ideal binary
collisions, under open boundary conditions and constant wind
(Fig.\ref{fig3}a), extracting a simple collision rule, i.e.~a
phenomenological function that relates the relative dune size $r_f$
after the collision between two dunes, and the corresponding initial
relative size $r_i$ and offset $\theta_i$ \cite{duran_condmat}. In accordance to recent underwater experiments \cite{hersen05} and our simulations, for most
initial conditions the number of dunes is conserved and the total sand volume
change is negligible.
Furthermore, as seen in Fig.~\ref{fig3}b, in most of the cases the collision increases the relative dune size ($r_f>r_i$), redistributing sand from large to small dunes, in sharp contrast to the flux balance, which accumulates sand on large dunes \cite{duran_condmat}.

\begin{figure*}[htb]
\begin{center}
\includegraphics[width=0.65\textwidth]{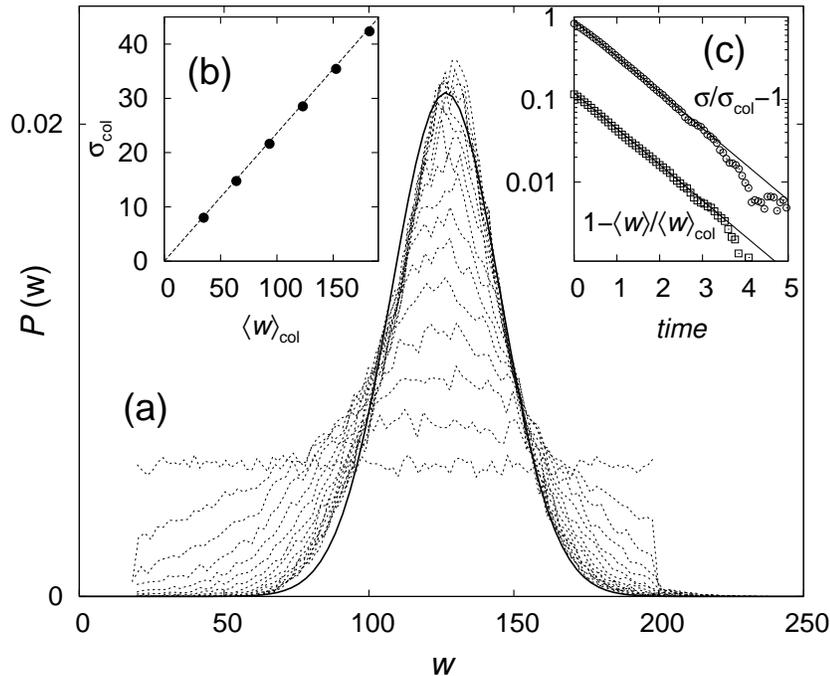}
\end{center}
\caption{\protect
{\bf (a)} Snap-shots of the evolution due to binary collisions of a uniform dune size distribution (dashed-lines) towards a Gaussian (solid line). {\bf (b)} Linear relation between the two first moments of the observed Gaussian, for different initial volumes of sand available, yielding a constant relative standard deviation $\sigma_{col}/\langle w \rangle_{col}$. {\bf (c)} The first two moments $\langle w\rangle$ (squares) and $\sigma$ (circles), exponentially relax toward their equilibrium values $\langle w\rangle_{col}$ and $\sigma_{col}$ (see text). Time units are number of collisions per dune.
}
\label{fig4}
\end{figure*}

With this collision rule we study the influence of collisions on the dune size distribution neglecting sand exchange through sand flux between them. A large sampling statistics of about 10 000 dunes is considered, where we assume that each pair of dunes collides with the same fixed probability following the collision rule introduced above (Fig.~\ref{fig3}b). This assumption does not consider the spatial distribution of dunes and thus we neglect the spatial correlations between dune sizes and positions. As a result of binary collisions, arbitrary initial size distributions  converge to a stationary Gaussian distribution $P_{col}(w)$, illustrated in Fig.~\ref{fig4}a. We find that the relative standard deviation $\sigma_{col}/\langle w\rangle_{col}$ of such distribution, where $\langle w\rangle_{col}$ is the mean size, is constant for different initial conditions (Fig.~\ref{fig4}b) and its value is fixed by the parameters of the dune model, which were chosen to reproduce the morphology of Moroccan dunes \cite{duran05,duran06b}, which in turn determine the dune morphology and thus the phenomenological collision rule \cite{duran_condmat}. The constant value of the relative standard deviation is consequence of the scale invariance of our collision dynamics that depends on the relative dune size instead of the absolute size. On the contrary,
with
changing wind conditions, the dune scale may influence the collision outcome and also several dunes may emerge from a single collision \cite{elbelrhiti08}.

Until now, two preliminary conclusions can be stated. First, the measured barchan dunes are approximately uniformly
placed over the deserts,
with
characteristic
values of $L$, and the different dune fields follow a common size distribution with a simple scaling with the mean dune size.
Second, neither sand flux alone does contribute to the size selection~\cite{hersen05,hersen04b} nor collisions alone can be responsible for the skewed measured non-Gaussian distributions (Fig.\ref{fig1}).
What results from the interplay between both mechanisms will be now derived
from a mean-field approach, leading
to a master equation for the size distribution $P(w,t)$ and
to a simple scaling between all the
three properties, $\langle w\rangle$, $S$ and $\langle L\rangle$.

\section{Master equation}
\label{sec:3}

When only ideal collisions occur, Fig.~\ref{fig4}c
shows that both, the mean $\langle w\rangle$ and the standard deviation
$\sigma$ of the size distribution exponentially relax toward their
respective equilibrium values, $\langle w
\rangle_{col}$ and $\sigma_{col}$. By using the definition
$\langle w\rangle(t)\equiv \int dw~w~P(w,t)$ and the exponential relaxation
$d\langle w\rangle/dt = (\langle w\rangle_{col}-\langle w\rangle)/t_c$,
the size distribution $P(w,t)$ obeys in first approximation the dynamical equation $dP/dt = (P_{col} - P)/t_c$, where $t_c$ is the characteristic relaxation time (in units of number of collisions per dune) of $P(w)$ towards the equilibrium Gaussian distribution $P_{col}$ when only collisions are considered.

When both collisions and sand flux balance are considered, the total temporal derivative of $P(w,t)$ has now two separated terms, the partial temporal derivative $\partial P/\partial t$ and the term
arising from the volume change rate $\dot{V}=\dot{w} dV/dw$,
due to flux balance: $\dot{V}\partial P/\partial V = \dot{w}\partial
P/\partial w$. Assuming the existence of a steady state and using
the empirical fact that $V\propto w^3$ and $\dot V\propto Q w$, with $Q$ denoting the saturated sand flux over a flat bed~\cite{elbelrhiti06}, the
master-equation yields
\begin{equation}
\frac{t_c}{t_s}\frac{\partial P(w)}{\partial w} =
             \frac{w}{\langle w\rangle^2_{col}}
             \left ( P_{col}(w)-P(w) \right ) ,
\label{boltzmann}
\end{equation}
where there are three parameters,
namely the relative deviation $\sigma_{col}/\langle w\rangle_{col}$ of $P_{col}$, determined by the collision model, the characteristic size $\langle w\rangle_{col}$ and the ratio $t_c/t_s$. Time $t_s$ is the characteristic time associated to the change rate of the dune size due to the sand flux balance, defined as $t_s = \alpha\langle w\rangle^2_{col}/Q$ with $\alpha$ as a constant. From Eq.~(\ref{boltzmann}) one concludes that when collisions dominate in the selection of dune sizes, $t_c \ll t_s$ and consequently the distribution converges to the Gaussian $P_{col}$. When the opposite occurs, with the sand flux balance being the relevant process, $P(w)$ deviates from $P_{col}(w)$ the more the larger $t_c/t_s$ is.

The solution of Eq.~(\ref{boltzmann}) is plotted in Fig.~\ref{fig1}f and
\ref{fig1}g (solid lines) for each dune field, with $\langle w\rangle_{col}$
and $t_c/t_s$ as fit parameters.
The value of $\sigma_{col}$ is taken from Fig.~\ref{fig4}b.
Apart extreme points, the solution fits well the empirical
distributions, with first and second moments reasonable approximated
as
\begin{subequations}
\begin{eqnarray}
\label{wtau}
\langle w \rangle & \simeq & 0.8 \langle w\rangle_{col} (\sqrt{t_c/t_s} + 1)\\
\label{stau}
S/\langle w\rangle  & \simeq & 0.62 \sqrt{t_c/t_s}/ (\sqrt{t_c/t_s} + 1)
\end{eqnarray}
\label{wtau_stau}
\end{subequations}
in the range $t_s < 5 t_c$.
From Eqs.~(\ref{wtau_stau}) one sees that the characteristic size $\langle w\rangle_{col}$ determined by the collisions dynamics is in fact the only characteristic size in the system. Moreover, Eq.~(\ref{stau}) shows that the relative standard deviation $S/\langle w\rangle$ is given by the ratio $t_c/t_s$ and thus describes a measure for the competition between sand flux balance ($t_s$) and collision ($t_c$) processes for the dune size selection.
For instance, dune field in Fig.~\ref{fig1}b has a large value
$t_c/t_s = 4.3$, indicating that the sand flux balance is the most
important size selection process, while the dune field in Fig.~\ref{fig1}c
has $t_c/t_s = 1.7$ indicating more relevance from collision processes.

Furthermore, the ratio $t_c/t_s$ is not an independent parameter since $t_c$ must be proportional to the collision time $t_{col}$, defined as the average time for two dunes to collide. This collision time is determined as the quotient between the inter-dune spacing $\langle
L\rangle$ and the average relative velocity between two dunes, $\langle v\rangle_r \propto \int_0^\infty dw_1\int_{w_1}^\infty dw_2 P(w_1) P(w_2) (v(w_1) - v(w_2))$.
Since the dune velocity follows $v \propto Q/w$ and,
within some wide range of sizes, $P(w)$ can be well approximated by a
log-normal distribution (see Figs.~\ref{fig1}f and \ref{fig1}g),
one obtains $\langle v\rangle_r \propto \langle v\rangle \sigma_{\ell}$
with $\langle v\rangle \propto Q/\langle w\rangle$ the dune average velocity
and $\sigma_{\ell}$
the standard deviation of the log-normal distribution
(adimensional),
yielding $t_c \propto \frac{\langle L\rangle \langle w\rangle}{Q\sigma_{\ell}}$.
Substituting $t_c$ and Eq.~(\ref{wtau}) into
Eq.~(\ref{stau}) and taking the first-order approximation $\sigma_{\ell}\sim S/\langle w\rangle$, we arrive at
\begin{equation}
\left(S/\langle w\rangle\right)^3 \simeq a \langle L\rangle/\langle w\rangle .
\label{sl}
\end{equation}
As shown in Figs.~\ref{fig1}h-i and \ref{fig2}d,
where one empirically
obtains $\epsilon = (S/\langle w\rangle)^{-1} =
2.9\sqrt[3]{\langle w\rangle/\langle L\rangle}$,
the scaling in Eq.~(\ref{sl}) indeed describes the measurements,
with the constant $a = 2.9^{-3}=0.041$, independent of the model parameters.

\section{Conclusions}
\label{sec:4}

In summary,
due to the dynamical nature of barchan dunes, the size distribution
is intrinsically linked to the spatial distribution in such a way that sparse
fields have a broader size distribution, while dense ones have narrow
distributions.
We have shown that the relative dune size distributions of Moroccan dune
fields collapse into an unique distribution function, and that dunes are
uniformly distributed with a characteristic inter-dune spacing that obeys
a simple scaling law.
By using a master-equation approach with a simple collision rule,
we showed that the simplest processes behind the change of the dune
size occurring in dune fields, namely ideal binary collisions and flux
balance, are able to properly determine the size distribution function.
Which mechanisms are behind the local selection of the specific size scale of a dune field remains an open question, since it involves not only binary collisions and flux balance under a stationary wind, but detailed processes in real changing wind conditions, i.e. calving~\cite{elbelrhiti06,elbelrhiti08},
that can locally change the dune size and are not included in the analysis we have presented.

\begin{acknowledgements}
The authors thank Maria Haase for useful discussions.
This research was supported in part by the Max-Planck prize.

\end{acknowledgements}



\end{document}